• Article •

# Niobium telluride absorber for a mode-locked vector soliton fiber laser

XinXin Shang[1], NanNan Xu[1], Jia Guo[2], Shuo Sun[3], HuaNian Zhang[3], S. Wageh[4], Ahmed A. Al-Ghamdi[4], Han Zhang[2,*], and DengWang Li[1,*]

[1]*Shandong Province Key Laboratory of Medical Physics and Image Processing Technology, Shandong Provincial Key Laboratory of Optics and Photonic Device, School of Physics and Electronics, Shandong Normal University, Jinan 250014, China*
[2]*Institute of Microscale Optoelectronics, Collaborative Innovation Center for Optoelectronic Science & Technology, Key Laboratory of Optoelectronic Devices and Systems of Ministry of Education and Guangdong Province, College of Physics and Optoelectronic Engineering, Shenzhen Key Laboratory of Micro-Nano Photonic Information Technology, Guangdong Laboratory of Artificial Intelligence and Digital Economy (SZ), Shenzhen University, Shenzhen 518060, China*
[3]*School of Physics and Optoelectronic Engineering, Shandong University of Technology, Zibo 255049, China*
[4]*Department of Physics, Faculty of Science, King Abdulaziz University, Jeddah 21589, Saudi Arabia*

Niobium telluride (NbTe$_2$), an emerging transition metal dichalcogenide material, has been theoretically predicted with nonlinear absorption properties and excellent optical response, while deficient research about utilized in ultrafast photonics has been reported. In this work, a NbTe$_2$-based saturable absorber was applied in an Er-doped fiber (EDF) as a mode-locked device and vector soliton was obtained based on NbTe$_2$ for the first time. NbTe$_2$ - PVA film SA was successfully prepared by liquid-phase exfoliation and spin-coating method, whose modulation depth is up to 10.87 %. The nonlinear absorption coefficient of NbTe$_2$ SA film tested through the open-aperture Z-scan laser measurement is $0.62 \times 10^{-11}$ m/W. Conventional soliton with pulse duration of 858 fs was generated based on NbTe$_2$ SA, which was demonstrated to be a kind of polarization-locked vector soliton in the further investigation. Our experimental results fully suggest the nonlinear optical properties of NbTe$_2$ and broaden its ultrafast applications as ultrafast photonic device.

**niobium telluride, saturable absorber, mode-locked pulse, fiber laser**

## 1 Introduction

Soliton is a kind of stable localized wave packet that is widespread in diverse fields from molecular chains, Bose-Einstein condensates, spin waves to nonlinear optics, which has been extensively expanded in the last few decades [1-3]. The generation of soliton can be attributed to the balance between nonlinearity and diffraction in space [4-6]. Optical soliton was experimentally recorded in single-mode fiber by Mollenauer et al. in 1980 for the first time [7]. Due to the abundant nonlinear effects in optical fiber, ultrafast fiber laser has been a reliable and meaningful research platform to study various soliton phenomena [8]. As in which effects of gain, loss, and dispersion play important roles, the soliton formed in fiber lasers which is a typical nonlinear dissipative system is normally dissipative soliton [9,10]. Theoretically, the dynamics of dissipative soliton is usually governed by the Ginzburg-Landau equation (GLE) [1]. Up to now, a variety of soliton phenomena, such as the conventional soliton, bound state soliton, soliton bunches, dispersion-managed soliton, restless soliton and so on, have been demonstrated in ultrafast fiber lasers based on nonlinear polarization rotation (NPR) technique. However, due to the use of polarization-sensitive devices, solitons achieved by NPR don't have polarization dynamics. Actually, the vector characteristics of solitons formed in fiber should be discussed since that there always exists birefringent in fiber under the twists and pressures. The formation of vector soliton which is a class of solitons that have two polarization components mutually coupled can be attributed to the cross-polarization coupling [11]. Within ultrafast fiber lasers, there are some works on the theoretical prediction of vector solitons have been done. The formation of stationary bound vector solitons was predicated by Haelterman *et al.* in 1993 [12]. Yang *et al.* found that the relative phases and initial separation are both the key of solitons repulsion or attraction [13]. Experimentally, Tang et al. demonstrated the high-order polarization-locked vector conventional solitons in a mode-locked fiber laser with a semiconductor saturable absorber mirror (SESAM) as the mode-locked device, and proved that both the group velocities of two components and the locked phase velocities play important roles in the formation process of soliton [14]. Zhao *et al.* recorded the bunched restless vector solitons with a SESAM [15]. In normal dispersion region, not only polarization-locked vector dissipative soliton but also polarization-rotating vector dissipative soliton was demonstrated by Zhang et al. [16].

---

*Corresponding author (Han Zhang, email: hzhang@szu.edu.cn; DengWang Li, email: dengwang@sdnu.edu.cn)

In fact, finding an appropriate polarization-independent saturable absorber is the key to obtaining vector soliton in fiber lasers. In recent years, the rapid development of research about nanomaterials has greatly promoted the wide applications of saturable absorbers (SAs) based on nanomaterials in the field of passively mode-locked lasers, and also provided a new direction for the study of vector solitons in fiber lasers [17-21]. Various materials including two-dimensional (2D) materials, metal oxide nanoparticles, quantum dots and so on have been employed in mode-locked lasers for generating ultrashort pulses [22-28]. 2D materials have an atom-thick layered structure and the weak Van der waals force provides the connecting force between the layers. They have peculiar electronic, optical and mechanical properties due to their unique 2D structures. As a typical 2D material, graphene has ultrahigh carrier mobility and unique zero bandgap value which make it have a huge application potential in photoelectric devices and ultrafast optics. In 2009, a SA based on graphene was prepared by Bao et al. for the first time, and ultrashort pulse output was obtained in an Er-doped fiber laser (EDFL) [29]. Since then, inspired by graphene, other 2D materials with similar crystal and band structures including topological insulators (TIs), transition metal dichalcogenides (TMDs), 2D single element materials (Xenes), MXenes, black phosphorus (BP) and so on have been utilized as saturable absorbers within mode-locked lasers [30,31]. TIs such as $Bi_2Se_3$, $Bi_2Te_3$ and $Sb_2Te_3$ possessing a bulk state with small energy gap and a surface state without energy gap have unique optoelectronic and nonlinear optical properties. In 2012, Bernad *et al.* firstly reported the saturable absorption of $Bi_2Te_3$ at 1.5 µm [32]. The novel Xenes (e.g., Borene, Gallene, Silene, Antimonene, Tellurene) exhibit huge application potential in emerging technology and basic science fields due to their excellent chemical, electronics, physical, and optical properties. MXenes with a general formula of $M_{n+1}AX_n$ (M: Sc, Ti, Cr, V, Nb, Hf, Ta, etc.; A: C, N; X: O, F, OH, etc.; n=1, 2, or 3) have a huge broadband absorption which their interband transitions and plasmon resonance peaks cover the entire UV, visible, and NIR range. MXenes have a great future in ultrafast photonics. Recently, BP as a novel 2D material arouse the interest of researchers because of its applications in the near-infrared range. But the thermal stability and air stability limit the further application in ultrafast mode-locked lasers.

TMDs whose general formula is $MX_2$ (M is on behalf of transition metal elements; X represents chalcogens) have unique electrical and optical properties due to their layer-dependent bandgap property [33]. When the TMDs are transformed from multilayer to monolayer, their energy bandgap structure also changes from indirect bandgap to direct bandgap. Generally, monolayer TMD exhibits a X-M-X sandwich structure. Wang *et al.* studied the ultrafast saturable absorption property of $MoS_2$ nanosheets and proved that $MoS_2$ dispersions exhibited better SA response compered to graphene dispersions under the same excitation condition in 2013 [34]. Under this reference, Du *et al.* achieved the first example of $MoS_2$-based fiber photonic device and obtained the mode-locked operation at 1 µm region within a Yb-doped fiber laser [35]. In 2014, vertically stood $WS_2$ nanoplates were studied by Fu et al., the nonlinear optical properties of the $WS_2$ nanoplates were measured firstly, which demonstrated that just nonlinear saturable absorption with negligible nonlinear refraction exists in $WS_2$ nanoplates, and showed that $WS_2$ is a kind of promising 2D nanomaterials for saturable absorber [36]. And next year, Mao *et al.* reported that $WS_2$ nanosheets not only have excellent ultrafast nonlinear saturable absorption characteristic but also exhibit high-damage threshold, and demonstrated an ultrafast fiber laser based on $WS_2$ SA [37]. In 2015, Luo *et al.* found that few-layer $MoSe_2$ existed both two-photon absorption and nonlinear saturable absorption, and built a 1.56 µm mode-locked soliton fiber laser in which few-layer $MoSe_2$ nanosheets were used as saturable absorbers for the first time [38]. Up to now, lots of soliton phenomena in fiber lasers based on TMDs-based saturable absorbers have been demonstrated. However, in fiber lasers, the research on vector solitons based on TMDs is not comprehensive enough [39]. Niobium telluride ($NbTe_2$) is a type of TMDs that has a structure of peculiar distorted octahedral (1T), where Nb atoms are polymerized into period tripling metal chains [40]. Recently, Liu *et al.* used NbTe$_2$ nanosheets as novel 2D photothermals agent and nanocarriers for the application in biomedical, ultrathin 1T-phase NbTe$_2$ single-crystalline nanosheets have distinctive photothermal properties, excellent drug loading rate and security, which offer great possibilities in biomedical fields, for instance, photographic developer, photothermal therapeutics and tissue imaging [41]. However, few research about vector soliton output from ultrafast fiber lasers based on NbTe$_2$ has been reported.

To begin with, we experimentally studied the nonlinear optical properties of NbTe$_2$-based SA. And then we systematically researched the vector soliton operation and the vector characteristics using a NbTe$_2$-based EDFL. The acquired vector soliton has a 858-fs pulse width with a 3.29-nm optical spectrum bandwidth. With a 54-dB signal-to-noise ratio (SNR), the EDFL operates with good stability. Besides, the long-time stability of NbTe$_2$-based SA is tested, and the result indicates that our homemade NbTe$_2$-based SA can withstand long-time lighting. Our work shows that TMDs, especially the NbTe$_2$, have great potential to be studied as SAs for the research about vector soliton forming in fiber lasers.

## 2 Preparation and Characterization of the NbTe$_2$ nanosheets

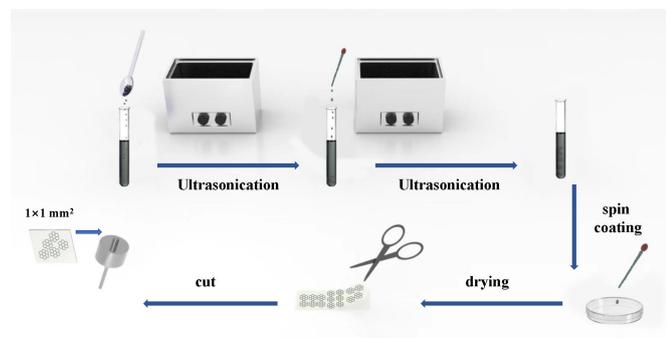

**Figure 1** The preparation process of the NbTe2-PVA SA

As shown in Figure 1, we show the preparation process of NbTe$_2$-PVA SA. At first, using liquid-phase exfoliation method, adding 50 mg NbTe$_2$ nanosheets into 50 mL alcohol to make dispersion solution. Set the blend in an ultrasonic cleaner for about 24 hours, and then centrifuged for 30 min with the speed of 1500 rpm so that we can prepare the layered nanosheets. Then, mixed 5 wt. % PVA solution and NbTe$_2$ solution which volume ratio is 1:1. Then cleaning mixture NbTe$_2$-PVA dispersion solution was attained by placed dispersion solution in an ultrasonic cleaner for about 8 hours. Next, spin coating method was used to get NbTe$_2$-PVA film. Spin-coating 100 μL NbTe$_2$-PVA dispersion solution on a petri dish to make NbTe$_2$-PVA film. Dry the dish in an oven for 24 h at 30℃. Lastly, cut off a thin film of $1 \times 1$ mm$^2$ from the petri dish. Put the film at the fiber end to be SA.

Chemical, structure and morphological properties are the basics that determine the application value of NbTe$_2$ in optoelectronic devices. So, we characterize the prepared NbTe$_2$ nanosheets before inserting the NbTe$_2$-based SA into an Er-doped fiber laser. The characteristics of NbTe$_2$ nanosheets are given in Figure 2. A scanning electron microscope (SEM) is used to test the surface morphology characteristics of the NbTe$_2$ nanosheets. The SEM image is shown in Figure 2(a), and distinct layered structure can be observed under a scale bar of 10 μm. Figure 2(b) illustrates the energy dispersive spectroscopy (EDS) pictures which indicate that the obtained NbTe$_2$ nanosheets only consists of Nb (35.26%) and Te (64.74%) without other elements. Further, an atomic force microscope (AFM) is adopted to test the thickness properties of the NbTe$_2$ nanosheets. Fig. 2(c) shows the test AFM pictures, in which the NbTe$_2$ nanosheets have a thickness of about 7-12 nm. Besides, a high-resolution transmission electron microscope (HRTEM) is used to test the lattice size of the NbTe$_2$ nanosheets under a resolution of 200 nm. And the HRTEM image is given in Figure 2(d), the lattice size is 3.68 Å which is conformed well with the published literature [42]. Figure 2(e) exhibits the Raman spectrum of NbTe$_2$, and there are three prominent peaks at around 93 cm$^{-1}$, 120 cm$^{-1}$, and 141 cm$^{-1}$ corresponding to the $A_g^2$, $B_g^3$, and $A_g^4$ modes, respectively [40]. The optical transmittance of NbTe$_2$ under different wavelengths can be obtained from Figure 2(f). And the transmissivity at 1550 nm is about 77.4%.

The open aperture Z-scan technique based on the beam spatial distortion is commonly used to measure the nonlinear absorption properties of materials. As shown in Figure 3(a), after a beam splitter the Gaussian beam emitted from an excitation light source is split into two beams. One of them shuts detector 1 directly, and another passes through a lens firstly and then is measured by detector 2 after passes through the sample needed to be tested. The zero position of the Z-axis is set on the focus of the lens and along the propagation direction of the beam is the positive direction of the Z-axis. Under the effects of self-focusing or self-defocusing effects of the sample, the intensity of the incident light will change as the sample moves in the positive direction of the Z-axis. The nonlinear absorption properties of the sample will be obtained by comparing two sets of data recorded by detectors 1 and 2. The excitation light source used is a pulsed laser whose operating wavelength is 1064 nm, and the repetition frequency and pulse width are 10 Hz and 25 ps. The waist radius at the zero position of the Z-axis is 42.4 μm. The normalized transmittance T(z) can be calculated by the equation as follows:

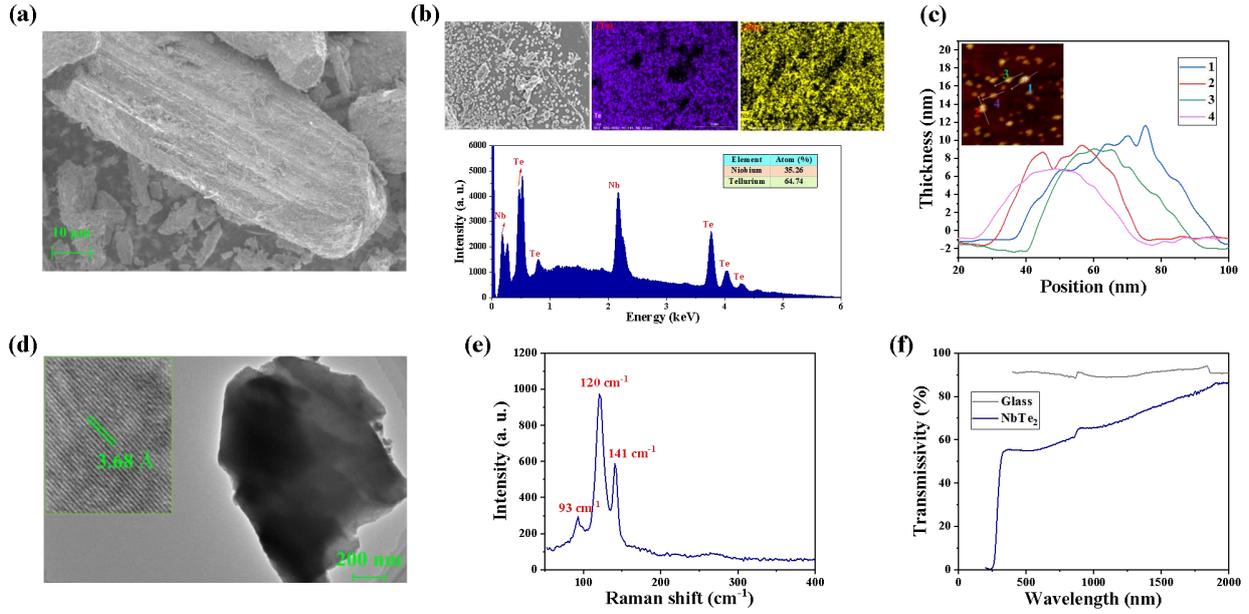

**Figure 2**  Characterizations of the NbTe₂ nanosheets. (a) The SEM image; (b) the EDS pictures; (c) the AFM pictures; (d) the TEM images and the HRTEM images; (e) the Raman spectrum; (f) transmittance versus wavelength.

$$T(z) = 1 - \beta I_0 L_{eff}/\left[2\sqrt{2}(1 + z^2/z_0^2)\right] \quad (1)$$

Where $\beta$ and $I_0$ represent the nonlinear absorption coefficient and the on-axis peak intensity at focus position (z=0), respectively. $L_{eff}$ is the effective length and $z_0$ stands for the Rayleigh range. As displayed in Figure 3(b), the nonlinear absorption coefficient ($\beta$) was fitted to be 0.62 × 10⁻¹¹ m/W. Besides, a balanced twin-detector system is constructed to measure the nonlinear saturable absorption of the NbTe₂-SA. Figure 3(c) exhibits schematic diagram of the balanced twin-detector measurement setup, a homemade ultrafast fiber laser (1564.8 nm, 353 fs, 10 MHz) is employed as a laser source, a variable optical attenuator (VOA) is adopted to adjust the input laser power and separated in a ratio of 1:1 by an OC, the power at both ends is tested by power meter. The obtained experimental datas are fitted by equation (2) as follows:

$$T(I) = 1 - T_{ns} - \Delta \cdot \exp\left(-\frac{I}{I_{sat}}\right) \quad (2)$$

Here, T(I) is the transmission velocity, $\Delta$ represents the modulation depth. I stands for the incident light intensity, $I_{sat}$ and $T_{ns}$ represent the saturation intensity and non-saturable absorbance, respectively. The experimental fitting results are given in Figure 3(d), the high saturation intensity is 19.65 MW/cm² and the modulation depth is calculated to be 10.87 %.

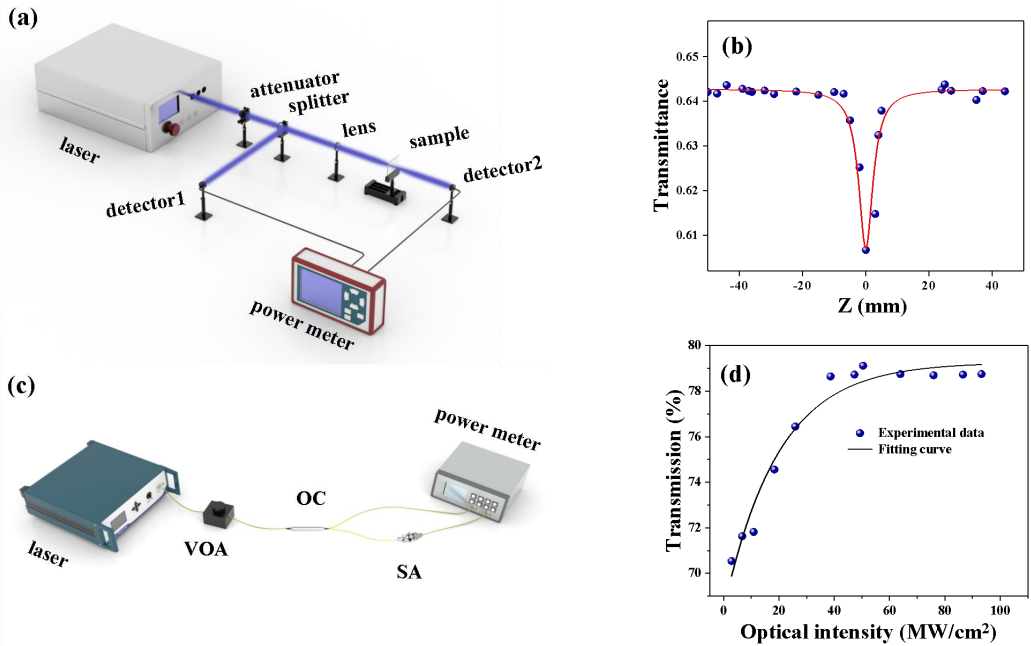

**Figure 3**  (a) Schematic diagram of the open aperture Z-scan experimental setup, (b) Z-scan fitting curve of the NbTe₂-PVA film, (c) schematic of the balanced twin-detector measurement setup, (d) the nonlinear transmission of NbTe₂-PVA SA.

## 3 Experimental setup

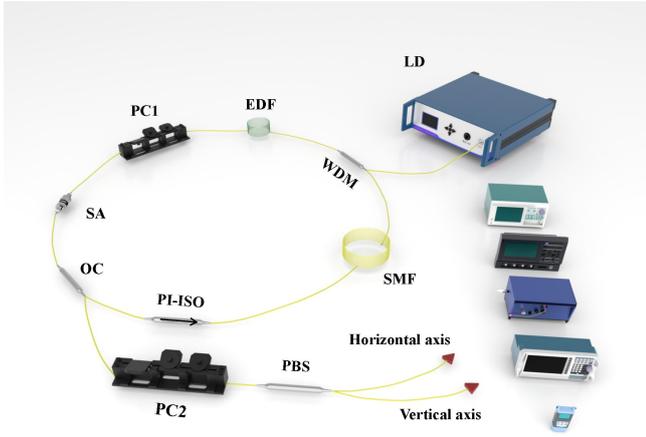

**Figure 4**  A schematic of the NbTe$_2$-based fiber laser

Figure 4 shows the schematic of the mode-locked fiber laser based on the NbTe$_2$-based saturable absorber which is constructed by sandwiching a NbTe$_2$-PVA film between two fiber ends and it is used as a mode-locker. A section of 10-m-long Er-doped fiber (EDF, OFS MP980, -18 ps/(nm·km)) is used as the gain fiber. A 976 nm laser diode (LD) is adopted to pump the gain fiber and input through a 980/1550 nm wavelength division multiplexer (WDM). The operating direction of the light in the ring cavity is controlled by a polarization-insensitive isolator (PI-ISO). A polarization controller (PC) is inserted into the cavity to adjust the intra-cavity linear birefringence in order to optimize the condition of mode-locking. The monitoring of the operating state of the fiber laser is achieved by the 30%-output part of an output coupler (OC). The beam is separated to two orthogonal polarization components by an in-line polarization beam splitter (PBS), and the polarization characteristics of the soliton are measured simultaneously. In order to avoid the fiber segment between the OC and the measuring device causing fiber birefringence, we inserted another PC between the OC and the PBS. Characteristics of the output pulses are tested by a 3 GHz photo-detector, a digital oscilloscope (Wavesurfer 3054), a radio-frequency (RF) spectrum (R&S FPC1000), an optical spectrum analyzer (AQ6317B, Yokogawa), and a power meter (PM3, Molectron, Barrington). The total cavity length is ~30.7 m with a 10-m-long EDF and all other fiber is standard single-mode fiber (SMF, 17 ps/(nm·km)), so the net dispersion of the cavity is calculated to be ~-0.22 ps$^2$.

## 4 Experimental results and discussions

Stable mode-locked operation can be recorded when the pump power is increased to 45 mW. Such low threshold is mainly because of that the NbTe$_2$-based SA has low saturation intensity and the insert loss is low. As shown in Figure 5(a), the EDFL works at 1561.17 nm with obvious Kelly sidebands, which is a typical characteristic of

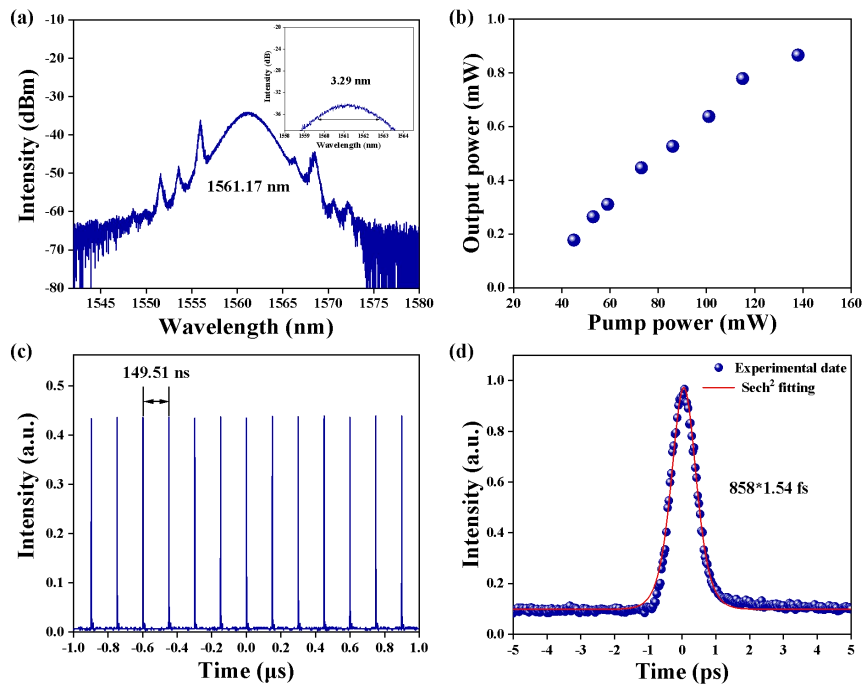

**Figure 5**  (a) Typical optical spectrum of conventional soliton; (b) variation of output power with pump power; (c) pulse trains; (d) autocorrelation trace of the conventional soliton.

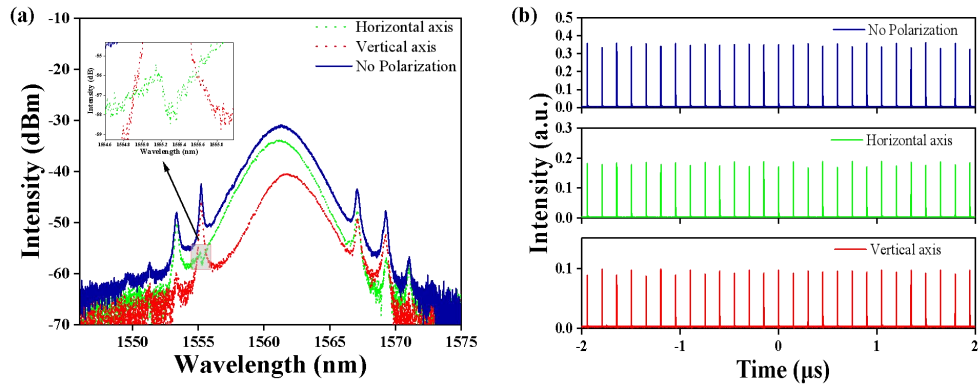

**Figure 6**  (a) Polarization-resolved spectra of the conventional soliton; (b) polarization-resolved pulse trains.

conventional soliton normally forming in a laser cavity with anomalous dispersion. Figure 5(b) gives the variation trend of output power, and with the pump power increasing from 45 mW to 138 mW, the output power changes from 0.18 mW to 0.86 mW. The recorded maximum single pulse energy is calculated to be ~0.13 nJ in this work. The output pulse trains with a pulse interval of ~149.51 ns are exhibited in Figure 5(c). The repetition rate is 6.68 MHz which corresponds well with the pulse interval, and the total cavity length can be calculated to be ~30.7 m. According to the autocorrelation trace shown in Figure 5(d), the pulse duration can be fitted to be ~858 fs by the $sech^2$ profile, corresponding to the 3 dB bandwidth of the optical spectrum of ~3.29 nm. The time-bandwidth product (TBP) is calculated to be ~0.352 under this condition. This value is a little larger than the transform limit value which is 0.315, and the result indicates the existence of a slight chirp in the fiber laser.

The vector nature of the soliton is studied as there is no polarization-sensitive component applied in our laser cavity. So a polarization beam splitter is utilized and connected with the 30% output part of the OC to test the vector characteristics of the conventional soliton. Figure 6(a) shows the polarization-resolved optical spectra of the soliton, including the spectra of no polarization, vertical axis and horizontal axis. From the partially enlarged picture in Figure 6(a), a spectral dip can be clearly observed from the horizontal component while the vertical component shows an obvious spectral peak. This difference between the two optical spectra is caused by the coherent energy exchange between the horizontal component and the vertical component which have different polarization states. Three polarization-resolved pulses trains that correspond to the polarization-resolved optical spectra in Figure 6(a) are given in Figure 6(b). Uniform pulse intensity and no existence of polarization rotation indicate that the obtained solitons are polarization-locked vector solitons. There is no linearly polarized soliton observed, although a coherent energy exchange is recorded in this work.

In order to test whether the $NbTe_2$-based SA can withstand long-time lighting, the long-time stability of optical spectrum is studied. As shown in Figure 7(a), the optical spectrum is recorded every hour for 12 hours, there is no obvious change observed, which indicates that the $NbTe_2$-based SA shows excellent long-term stability. Besides, the RF spectrum is shown in Figure 7(b), the fundamental repetition rate of the conventional soliton is 6.68 MHz, corresponding to the SNR of ~54 dB, presenting that the EDFL operating with low-amplitude fluctuation and good stability. What's more, the RF spectrum with a wide band of 200 MHz is also tested and given in Figure 7(c), which indicates that the higher-order harmonic waves exist good stability too.

## 5  Conclusions

In conclusion, we experimentally studied the nonlinear optical properties of $NbTe_2$-based SA and systematically

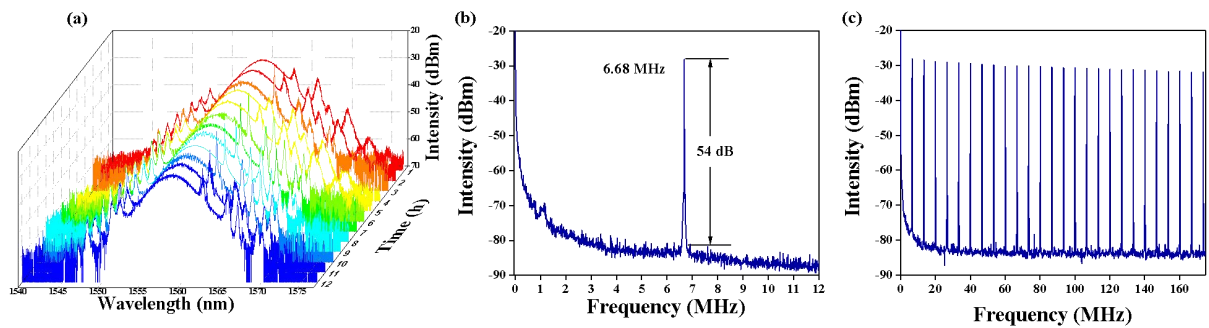

**Figure 7**  (a) Time stability of the optical spectrum; (b) frequency spectrum; (c) RF spectrum with a bandwidth of 200 MHz.

researched the vector soliton operation and the vector characteristics using a NbTe$_2$-based EDFL. The acquired vector soliton has a 858-fs pulse width with a 3.29-nm optical spectrum bandwidth. With a 54-dB SNR, the EDFL operates with good stability. Besides, the long-time stability of NbTe$_2$-based SA is tested, and the result indicates that our homemade NbTe$_2$-based SA can withstand long-time lighting. Our work shows that TMDs, especially the NbTe$_2$, have great potential to be studied as SAs for the research about vector soliton forming in fiber lasers.

*This work was supported by the National Natural Science Foundation of China (Grant nos. 61971271，11904213, 11747149), the Jinan City-School Integration Development Strategy Project (JNSX2021023)，and the Shandong Province Major Technological Innovation Project (2022CXGC010502), Shandong Province Natural Science Foundation (ZR2018QF006), the Key Project of Department of Education of Guangdong Province (Grant no.2018KCXTD026), the Deanship of Scientific Research (DSR) at King Abdulaziz University, Jeddah (Grant no: KEP-MSc-70-130-42), supported by "Opening Foundation of Shandong Provincial Key Laboratory of Laser Technology and Application".*